\documentclass[aip,amsmath,amssymb,reprint,floatfix
]{revtex4-1}
\usepackage{hyperref}
\usepackage{verbatim}
\usepackage{epsfig,graphicx}
\usepackage[english]{babel}

\makeatletter

\newcommand\gsl{\ifmmode\textsl{g}\else g\fi}

\makeatother

\begin{document}

\preprint{AIP/123-QED}

\title{Dynamics of horizontal buoyant jets.}

\author{V.~P.~Goncharov}
\email{v.goncharov@rambler.ru}
\affiliation{A. M. Obukhov Institute of Atmospheric Physics RAS, 109017 Moscow, Russia}

\date{\today}

\begin{abstract}
A single-layer model is used to construct the theory of straight-line jets. In addition to stationary states, the evolution of such jets admits transverse pulsations. The theory predicts that the period for warm jets pulsations is longer than the period of inertial oscillations caused by the Earth's rotation, while for cold jets pulsations, it is shorter. Thus, only warm jets can have a noticeable effect on the synoptic range of the wind-speed spectrum, generating into it additional spectral peaks. Within such an approach, every a single jet is described by a self-similar compactly localized solution that can be considered as a constant-shear band.
\end{abstract}

\maketitle

\section{Introduction}

Jet streams in the Earth's atmosphere are fast flowing, narrow air currents produced by two factors: the atmospheric heating and the influence of the Coriolis force caused by the planet rotation~\citep{rhi94,val06,hh13}. Being one of the elements in the atmospheric dynamics, jet streams can be responsible for strong long-period fluctuations of the horizontal wind. In particular, the presence of additional peaks in the synoptic range of the spectrum of horizontal wind speed we can consider as a confirmation of this effect.

\begin{figure}[t!]
\centerline{\includegraphics[width=\columnwidth]{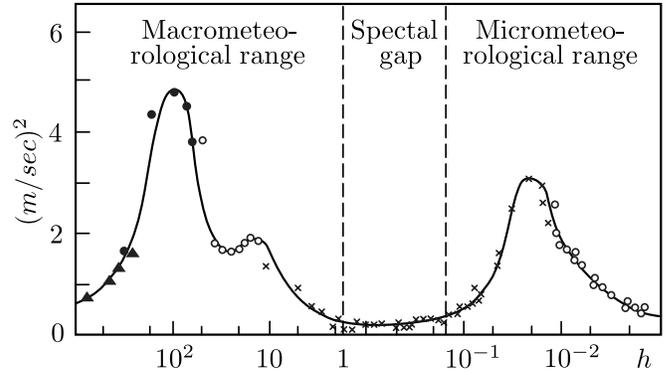}}
\caption{Power spectral density of horizontal wind speed by Van der Hoven~\citep{vdh57}.}\label{fig1}
\end{figure}

The main spectral properties of the horizontal wind velocity established by Van der Hoven~\citep{vdh57} are shown in Fig.~\ref{fig1}. This spectrum has three well-defined peaks. One of them called the synoptic maximum is at the period about $4\,days$. The next peak corresponds to half a day, which is close to the period of inertial oscillations caused by the Earth's rotation. And the last known as the microscale maximum locates at a period of $1\,min$. The frequency range between the synoptic and microscale peaks contains almost no energy and is known as the spectral gap.

Beginning with the classic work of Van der Hoven~\citep{vdh57} and up to the present day, there has been accumulating enough experimental evidence that the macro-meteorological spectral range may also include spectral peaks for fluctuations with periods exceeding inertial. Examples of such observations are presented in Kang and Won~\citep{kw15,kw16}(see also the references quoted therein).

A reasonable explanation of additional spectral peaks and the macro-meteorological spectral maximum can be obtained within shallow water theory, assuming that the jets are horizontal narrow flows of one fluid at the bottom of another and evolve under the action of gravity and the Coriolis force created by the rotation of the Earth.

Depending on the ratio between the fluid density inside such a jet and the background density of fluid around it,  its evolution has three alternative scenarios. The joint action of the Coriolis and the Archimedean (buoyancy) forces either causes transverse pulsations of this jet, or leads to its narrowing and increasing in height, or makes it monotonically expand with decreasing in altitude.

It is noteworthy that similar behavior resulting from effects of the buoyancy and rotation is detected for vortex structures considered within the active layer model~\citep{del03,gp12,gp13a,gp13b,gp15,gp16}. These structures look like drops or fingers that are inside a homogeneous fluid and have elliptic contact with a horizontal, rigid bottom. Being by solutions of a conservative dynamical system, they can rotate and change their semi-axes, and hence, the contact area and the height without changing the volume, total angular momentum as well as other integrals of motion.

The drop- or finger-like structures are self-similar solutions and inherent to many of the so-called active layer models, which describe a depth-averaged flow of incompressible fluid with a horizontally-nonuniform density in the lower (active) layer. Application of such models provides a way to understand the different real processes initiated by the Rayleigh-Taylor instability. In addition, since these structures produce space-time singularities responsible for the power-law tails in the short-wave range of the spectrum, their study provides a key to the understanding of strong turbulence~\citep{knnr07}.

As practice has shown, using upward jets as structural elements of the theory, these models can quite adequately describe the vertical mixing, transport processes, and turbulence in astrophysical and geophysical flows~\citep{del03,gp12,gp13a,gp13b,gp15,gp16}. The results obtained in these works inspire a hope that horizontal jet currents can also get a description within the active layer models. One argument in favor of this is that, in the limiting case, when one of the ellipse semi-axes in the base of drop tends to zero, the vortex drop-like structures in these models can transform into infinitely thin horizontal jets.

The generation of oscillations by horizontal jets due to their transverse pulsations deserves a more detailed discussion and will be presented in the next sections. Here, we note only that pulsational periods of warm jets should be longer than the period of inertial oscillations and therefore lie in the synoptic range. This fact can be argued with purely qualitative considerations.

Let's imagine that a fluid fragment has a density contrast with ambient fluid and is under the influence of gravity and the Coriolis forces. If the rotation is absent, the action of gravity leads to only two regimes. The fluid fragment either sinks to the bottom or rises upwards, depending on whether or not it is heavier or lighter than the ambient fluid. Because of irreversibility, both effects can be treated as instability. And conversely, if we eliminate gravity, then the action of the Coriolis force can cause only oscillations. Thus, we can conclude that both the instabilities and oscillations are regimes equally accessible for the active layer models.

According to the theory of dynamic systems, such symbiosis of regimes implies a specific topology of dynamical trajectories in the phase space of the system. In this space, stable regimes corresponding to oscillations evolve along closed trajectories, while the unstable along the non-closed. In doing so, the zones of stability and instability separate each other by closed loops known as separatrices. It is due to the active layer models have separatrices the oscillation periods caused by gravity and by the Earth’s rotation can undergo significant changes in response to a small increase in the amplitude of these oscillations.

Although the present paper limits to study only oscillatory regimes for horizontal jets, the Rayleigh-Taylor instability can also make a notable impact on the spectrum of wind speed fluctuations. The latter mechanism is of importance when studying large-scale turbulence, sometimes called a structural because its structural elements are narrow upward jets. In particular, a similar approach was used to explain a scaling effect in the spectra of the Sun’s supergranulation~\citep{gp14}. In this work, the supergranulation was considered as a collective effect produced by a statistical ensemble of upward jets.

This paper is structured as follows. In section~\ref{sec2}, we discuss the active layer model in the context of its following application to study horizontal jet streams. In section~\ref{sec3}, this model is used to derive the master equations for describing straight-line jets with transverse pulsations. In section~\ref{sec4}, we analyze self-similar solutions and give examples for jet profiles that could provide these kinds of solutions. The pulsational dynamics, impacting not only transverse parameters of jets but also components of their velocity, are discussed in section~\ref{sec5}. One of the goals of this section is to show that the period of inertial fluctuations resulting from the Coriolis effect is the shortest period for pulsations of warm jets. Possible nonlinear mechanisms to increase this period are also discussed here. In section~\ref{sec6}, we summarize our results. The appendix provides some details omitted in section~\ref{sec5} when integrating the nonlinear differential equation.

\section{Active layer model}\label{sec2}

We consider a two-dimensional field model whose evolution is governed by the equations
\begin{gather}
\frac{du}{dt}-fv=-\frac{1}{2}h\tau_{x}-\tau h_{x},\label{eq:1}\\
\frac{dv}{dt}+fu=-\frac{1}{2}h\tau _{y}-\tau h_{y},\label{eq:2}\\
\frac{dh}{dt}+h\left(u_{x}+v_{y}\right)=0,\quad\frac{d\tau}{dt}=0.\label{eq:3}
\end{gather}
The notation is as follows: $x$, $y$ are the Cartesian horizontal coordinates, $d/dt$ is the total temporal derivative, subscripts denote the partial derivatives, and $u$, $v$ are the depth-averaged components of the horizontal velocity.

Eqs.~(\ref{eq:1})--(\ref{eq:3}) describe the depth-averaged flow of incompressible fluid in an boundary layer of thickness $h(x,y,t)$ in shallow-water approximation, on assuming of hydrostatic balance and of rotation with the constant angular velocity $\frac{1}{2}f$ about the vertical axis. The members $fv$ and $fu$ on the left side of Eqs.~(\ref{eq:1}), (\ref{eq:2}) implies the components of the Coriolis acceleration. The physical specification of the variable $\tau$ depends on what assumption about the density distribution is used in the model.

If we suppose that the boundary layer rests on an impermeable horizontal bottom and is bounded above by a free surface~\citep{rip93}, then the variable $\tau$ has the meaning of the total buoyancy and is defined as
\begin{equation}
\tau=\gsl(1+\Delta\varrho/\varrho_{0}),\label{eq:4}
\end{equation}
where $\gsl$ is the gravity, and $\Delta\varrho$ is the density deviation from the background value $\varrho_{0}$. According to this definition, the total buoyancy $\tau$, along with the thickness $h$, is a strictly positive function of horizontal coordinates and time. The formulation based on Eqs.~(\ref{eq:1})--(\ref{eq:3}) with such definition of buoyancy is known as the thermal rotating shallow-water model (TRSW) (see Zeitlin~\citep{ztl18}, sec.~14 and references therein).

Note that in the case when $\Delta\varrho=0$, so that $\tau=\gsl$ is constant, Eqs.~(\ref{eq:1})--(\ref{eq:3}) reduce to the usual shallow water equations describing motion under the influence of gravity $\gsl$ and the Coriolis force.

In the atmosphere where the free boundary condition is hardly applicable, it is relevant to refuse it by transforming the TRSW model into the so-called active layer model~\citep{del03,gp12,gp13a,gp13b,gp15}. For this, all we need to do is add one more layer of an incompressible fluid with the density $\varrho_{0}$ so that it fills all the half-space above the bottom layer up to infinity.

If both of these fluids are under the action of gravity and Coriolis forces, and the horizontally-nonuniform density jump $\Delta\varrho$ between them is small, so that $\Delta\varrho/\varrho_{0}\ll1$, then in such models~\citep{del03,gp12,gp13a,gp13b,gp15} the shallow-water approximation, the depth-averaging, and the requirement of hydrostatic balance leads us again to Eqs.~(\ref{eq:1})--(\ref{eq:3}). But now, instead of (\ref{eq:4}), these models use the relation
\begin{equation}
\tau=\gsl\Delta\varrho/\varrho_{0},
\end{equation}
which gives any sign for $\tau$. Thus, in the active layer model, the variable $\tau$ has the meaning of the relative buoyancy, rather than merely buoyancy as in the condition (\ref{eq:4}), where it takes only positive values.

When density variations are induced only by temperature ones $\Delta T$ and are linearly connected, the relative buoyancy can be computed as
\begin{equation}
\tau=-\gsl\Delta T/T_{K},\label{eq:5}
\end{equation}
where $T_{K}$ is the background absolute (Kelvin) temperature. This parametrization allows taking into account both heating and cooling effects~(see as examples~\citep{and84,rip99}).

As with any non-dissipative system, the active-layer model has symmetries which imply the existence of conservation laws. Among them are the total energy
\begin{equation}
H=\frac{1}{2}\int h\left(u^{2}+v^{2}+h\tau\right)dxdy,\label{eq:6}
\end{equation}
the total angular momentum $S=M+fI/2$, and the quantity
\begin{equation}
G=\frac{1}{4}\left(\frac{dI}{dt}\right)^{2}+\left(M+2f^{-1}H\right)^{2}\label{eq:7}
\end{equation}
that remains unchanged due to the scaling invariance~\citep{gp14}. Here the angular momentum $M$ and the moment of inertia $I$ are described by the expressions:
\begin{equation}
M=\int h(xv-yu)dxdy,\quad I=\int h\left(x^{2}+y^{2}\right)dxdy.\label{eq:8}
\end{equation}

Additionally, the model leaves invariant the integral
\begin{equation}
C=\int h\left(F_{1}(\tau)+qF_{2}(\tau)\right)dxdy,\label{eq:9}
\end{equation}
where $F_{1,2}$ are arbitrary functions of the buoyancy $\tau$, and $q$ is the potential vorticity, which is defined as
\begin{equation}
q=h^{-1}\left(f+v_{x}-u_{y}\right),\label{eq:10}
\end{equation}
and obeys the equation
\begin{equation}
\frac{dq}{dt}=\frac{1}{2h}\left(h_{x}\tau_{y}-h_{y}\tau_{x}\right).\label{eq:11}
\end{equation}

\section{Straight-line jets}\label{sec3}

\begin{figure}[t!]
\centerline{\includegraphics[width=\columnwidth]{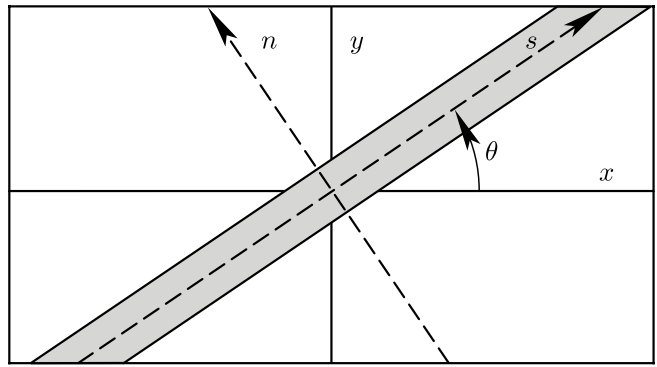}}
\caption{Scheme of a straight-line jet flow, which is oriented along the axis $s$ and pulsates transversely to it.}\label{fig9}
\end{figure}

Next, the active layer model will be applied to study so-called straight-line jets on the assumption that their dynamics includes only the changes in direction and transverse pulsations as shown in Fig.~\ref{fig9}. To formulate the motion equations for these jets in a more easy-to-use form, we introduce, instead of $x$ and $y$, new coordinates
\begin{equation}
s=x\cos\theta+y\sin\theta,\quad n=y\cos\theta-x\sin\theta.\label{eq:12}
\end{equation}
Here $s$ and $n$ are the longitudinal and transverse coordinates accompanying the jet's angular motion, and $\theta(t)$ is the angle of its slope relative to $x$-axis.

In the coordinates $s$, $n$, new velocities are defined as
\begin{equation}
U=\frac{ds}{dt},\quad V=\frac{dn}{dt},\label{eq:13}
\end{equation}
and are connected with the old ones by the relations
\begin{gather}
u=U\cos\theta-V\sin\theta-
\left(s\sin\theta+n\cos\theta\right)\theta_{t},\label{eq:14}\\
v=U\sin\theta+V\cos\theta +\left(s\cos\theta-n\sin\theta\right)\theta_{t}.\label{eq:15}
\end{gather}

Taking into account these formulas, after some simplifications, we find that in the new coordinates the equations of motion (\ref{eq:1})--(\ref{eq:3}) take the form
\begin{gather}
\frac{dU}{dt}-2V\theta_{t}-s\theta_{t}^{2}-n\theta_{tt}-
f\left(V+s\theta_{t}\right)=-h_{s}\tau-\frac{1}{2}h\tau_{s},\label{eq:16}\\
\frac{dV}{dt}+2U\theta_{t}-n\theta_{t}^{2}+s\theta_{tt}+
f\left(U-n\theta_{t}\right)=-h_{n}\tau-\frac{1}{2}h\tau_{n},\label{eq:17}\\
h_{t}+\partial_{s}\left(Uh\right)+\partial_{n}\left(Vh\right)=0,\label{eq:18}\\
\tau_{t}+U\tau_{s}+V\tau_{n}=0.\label{eq:19}
\end{gather}

We are looking for solutions that can be physically interpreted as homogeneous (along the $s$-axis) jets whose dynamics and therefore the variables $h$, $\tau$, $U$, $V$ depend only on time and the transverse coordinate $n$. As the analysis shows, for such jet-currents, the angular velocity $\theta_{t}$ must either be absent $(\theta_{t}=0)$ or constant $(\theta_{t}=-f)$. For both these cases the equations of motion (\ref{eq:16})--(\ref{eq:19}) can be rearranged to the form
\begin{gather}
\tau_{t}+V\tau_{n}=0,\label{eq:20}\\
U_{t}+VU_{n}-\omega V=0,\label{eq:21}\\ h_{t}+\partial_{n}\left(Vh\right)=0,\label{eq:22}\\
V_{t}+VV_{n}+\omega U=-h_{n}\tau-\frac{1}{2}h\tau_{n},\label{eq:23}
\end{gather}
where the parameter $\omega$ takes two values: $\omega=f$ if $\theta_{t}=0$, and $\omega=-f$ if $\theta_{t}=-f$.

It is easy to verify that, as before, Eqs.~(\ref{eq:20})--(\ref{eq:23}) conserve the integrals (\ref{eq:6}) and (\ref{eq:9}), but only in a one-dimensional form
\begin{gather}
H=\frac{1}{2}\int h\left(U^{2}+V^{2}+h\tau\right)dn,\label{eq:24}\\
C=\int h\bigl(F_{1}\left(\tau\right)+
qF_{1}\left(\tau\right)\bigr)dn, \label{eq:25}
\end{gather}
where now the role of the potential vorticity $q$ is played by the quantity
\begin{equation}
q=h^{-1}\left(\omega-U_{n}\right),\label{eq:26}
\end{equation}
which according to (\ref{eq:21}), (\ref{eq:22}) obeys the equation
\begin{equation}
\frac{dq}{dt}=q_{t}+Vq_{n}=0.\label{eq:27}
\end{equation}

As for the rest of the laws of conservation, they turn out to be unavailable in the one-dimensional version because reducing the dimensionality leads to the loss of certain symmetries.

\section{Self-similar solutions}\label{sec4}

By limiting to the class of self-similar solutions, when examining  Eqs.~(\ref{eq:20})--(\ref{eq:23}) we will assume that the buoyancy $\tau$ depends on $n$ and $t$ in a self-similar manner as
\begin{equation}
\tau=\hat{\tau}(z),\quad z=n\beta^{-1},\label{eq:28}
\end{equation}
thus the new independent variable $z$ is the self-similarity variable, and $\beta=\beta(t)$ is a function of time.

The assumption of self-similarity allows us to make a few simplifications. Indeed, by using the ansatz (\ref{eq:28}) in Eq.~(\ref{eq:20}), we at once obtain the expression for the transverse velocity
\begin{equation}
V=-\tau_{t}/\tau_{n}=\beta_{t}z.\label{eq:29}
\end{equation}

Substitution of this result into Eqs.~(\ref{eq:21}), (\ref{eq:22}) gives us three more constraints
\begin{equation}
h=\beta^{-1}\hat{h}(z),\quad U=\omega\alpha z,\quad\alpha_{t}=\beta_{t},\label{eq:30}
\end{equation}
where $\hat{h}(z)$ is some function which along with $\hat{\tau}(z)$ is to be found. In besides, the condition imposed on the longitudinal velocity $U$ allows us to interpret the obtained solution as a shear band, which from outside looks like a flow composing of two counter jets.

Finally, utilizing the obtained results (\ref{eq:29}), (\ref{eq:30}) and applying the method of separation of variables, from Eq.~(\ref{eq:23}) it follows that, on the one hand,
\begin{equation}
\beta_{tt}+f^{2}\alpha-\kappa\beta^{-2}=0,\label{eq:31}
\end{equation}
and on the other hand, the profile functions $\hat{h}(z)$ and $\hat{\tau}(z)$ obey the equation
\begin{equation}
\partial_{z}\bigl(\hat{h}^{2}\hat{\tau}\bigr)+2\kappa z\hat{h}=0.\label{eq:32}
\end{equation}
Here $\kappa$ is an arbitrary constant known as the separation constant.

Since the velocities $V$ and $U$ depend linearly on z, to be physically reasonable, the solutions of Eq.~(\ref{eq:32}) must be entirely localized on a finite interval of $z$ and vanishing outside of it. Besides, as the quantity $h$ is strictly positive, due to the invariance with respect to the replacement $z\rightarrow-z$, Eq.~(\ref{eq:32}) can have only even solutions. Therefore, without loss of generality, we may seek its solutions on the interval $-1\leq z\leq1$ in assuming they are symmetric about $z=0$ and subject to the boundary conditions
\begin{equation}
\hat{h}|_{z=\pm1}=0,\quad\hat{\tau}|_{z=\pm1}=0.\label{eq:33}
\end{equation}

The physical meaning of the separation constant $\kappa$ becomes clear if paying attention to the fact that it connects with two integrals of the problem
\begin{equation}
\mathfrak{U}=\frac{1}{2}\int h^{2}\tau dn,\quad I=\int n^{2}hdn.\label{eq:34}
\end{equation}
One of them $\mathfrak{U}$ is the potential energy, and the other $I$ is known as the total momentum inertia. In the self-similar case, both integrals can be presented as
\begin{gather}
\mathfrak{U}=\beta^{-1}\mathfrak{U}_{0},\quad I=\beta^{2}I_{0},\label{eq:35}\\
\mathfrak{U}_{0}=\frac{1}{2}\int_{-1}^{1}\hat{h}^{2}\hat{\tau}dz,\quad
I_{0}=\int_{-1}^{1}z^{2}\hat{h}dz,\label{eq:36}
\end{gather}
where $\mathfrak{U}_{0}$ and $I_{0}$ are constants functionally depended on profile functions $\hat{h}$ and $\hat{\tau}$. Multiplying Eq.~(\ref{eq:32}) on $z$ and integrating over the interval $-1\leq z\leq1$, one can show that
\begin{equation}
\kappa=\mathfrak{U}_{0}/I_{0}.\label{eq:37}
\end{equation}

Another property can be directly derived from Eq.~(\ref{eq:32}) if rewriting it as
\begin{equation}
\hat{\tau}=\frac{2\kappa}{\hat{h}^{2}}\int_{z}^{1}z\hat{h}dz,\label{eq:38}
\end{equation}
Thus, as follows from this relation, the signs of the separation constant $\kappa$ and the relative buoyancy $\hat{\tau}$ always coincide.

To construct solutions for the profile functions $\hat{h}$ and $\hat{\tau}$, we will consider a generating function $\psi(z)$, which has the following properties.
First, being defined in the interval $-1\leq z\leq1$, it changes the sign of its derivative $d\psi/dz$ in the point $z=0$, and second, near the points $z=\pm1$ it behaves as
\begin{equation}
\psi\approx\left(1-z^{2}\right)^{1+\gamma},\label{eq:39}
\end{equation}
where the parameter $\gamma$ satisfies the inequality $0\leq\gamma\leq1$.

Then for any such function $\psi(z)$, solutions of Eq.~(\ref{eq:32}) can be presented in the following form
\begin{equation}
\hat{h}=-\frac{1}{z}\frac{d\psi}{dz},\quad\hat{\tau}=
2\kappa\frac{\psi}{\hat{h}^{2}}.\label{eq:40}
\end{equation}

It is easy to check that under the condition (\ref{eq:39}), the solutions (\ref{eq:40}) possess all the necessary properties. They provide positivity of the function $\hat{h}$ and turn it together with $\hat{\tau}$ into zero at points $z=\pm1$ according to the power laws
\begin{equation}
\hat{h}\propto\left(1-z^{2}\right)^{\gamma},\quad
\hat{\tau}\propto\left(1-z^{2}\right)^{1-\gamma}.\label{eq:41}
\end{equation}

Despite the functional arbitrariness in the choice of the generating function $\psi$, all jet-streams with the same parameter $\kappa$, have the identical transverse dynamics. This fact follows immediately from Eq.~(\ref{eq:31}), since under fixed $\omega$ the evolution of the self-similarity variable $\beta(t)$ depends only on $\kappa$.

\begin{figure}[t!]
\centerline{\includegraphics[width=\columnwidth]{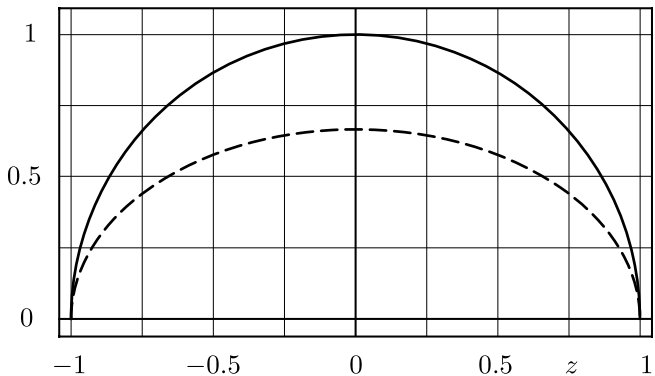}}
\caption{The normalized profiles in cross-section of the jet, which are generated by the function (\ref{eq:43}). The solid line shows the thickness $\hat{h}/a$ and the dashed line displays the buoyancy $\hat{\tau}a/\kappa$.}\label{fig2}
\end{figure}

\begin{figure}[t!]
\centerline{\includegraphics[width=\columnwidth]{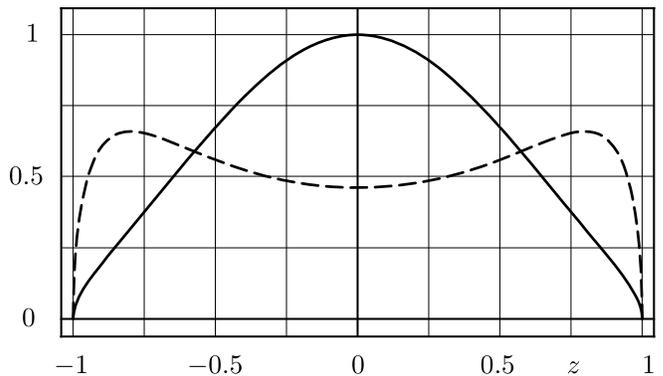}}
\caption{The normalized profiles in cross-section of the jet, which are generated by the function (\ref{eq:46}). The solid line shows the thickness $\hat{h}/a$ and the dashed line displays the buoyancy $\hat{\tau}a/\kappa$.}\label{fig3}
\end{figure}

Two pairs of the solutions with different profiles, but with the same parameter $\kappa$, are presented in Figs.~\ref{fig2},~\ref{fig3}.
The first of them includes the profiles
\begin{equation}
\hat{h}=a\left(1-z^{2}\right)^{1/2},\quad
\hat{\tau}=\frac{2\kappa}{3a}\left(1-z^{2}\right)^{1/2},\label{eq:42}
\end{equation}
corresponding to the generating function
\begin{equation}
\psi=\frac{a}{3}\left(1-z^{2}\right)^{3/2},\label{eq:43}
\end{equation}
where $a>0$ is some arbitrary constant.

The other pair consists of the profiles
\begin{gather}
\hat{h}=ae^{-z^2}\sqrt{1-z^2},\label{eq:44}\\
\hat{\tau}=\frac{\kappa}{a}\frac{e^{z^{2}}}
{\sqrt{1-z^{2}}}\left(1-\frac{D\left(\sqrt{1-z^{2}}\right)}
{\sqrt{1-z^{2}}}\right),\label{eq:45}
\end{gather}
generated by the function
\begin{equation}
\psi=\frac{a}{2}e^{-z^{2}}\left(\sqrt{1-z^{2}}-
D\left(\sqrt{1-z^{2}}\right)\right).\label{eq:46}
\end{equation}
where $D$ is the Dawson function.

\section{Pulsations of jets}\label{sec5}

Within the framework of this model, the pulsational dynamics of jet structures involves the changes not only in their transverse parameters but also in the longitudinal component of velocity. Completely presented by Eqs.~(\ref{eq:29})--(\ref{eq:31}), the jet dynamics is described by two variable $\alpha$ and $\beta$ which according to (\ref{eq:29}) are related as
\begin{equation}
\alpha=\beta-c,\label{eq:47}
\end{equation}
where $c$ is an integration constant.

The elimination of $\alpha$ allows us to formulate the motion equation (\ref{eq:31}) in the Hamiltonian form
\begin{equation}
\beta_{t}=H_{\lambda},\quad\lambda_{t}=-H_{\beta}\label{eq:48}
\end{equation}
with the Hamiltonian $H$, defined by the expression
\begin{equation}
H=\frac{1}{2}\left(\lambda^{2}+f^{2}\beta^{2}\right)+
\frac{\kappa}{\beta}-cf^{2}\beta.\label{eq:49}
\end{equation}

The constant $c$ can be fixed by assuming that the oscillatory system has a stable stationary state in which it obeys the conditions
\begin{equation}
H_{\lambda}=0,\quad H_{\beta}=0,\quad H_{\beta\beta}>0.\label{eq:50}
\end{equation}
Whence, supposing that this state is achieved at a value of $\beta=\beta_{0}$, we find the two constraints
\begin{equation}
c=\beta_{0}-\frac{\kappa}{f^{2}\beta_{0}^{2}},\quad
\frac{2\kappa}{f^{2}\beta_{0}^{3}}>-1.\label{eq:51}
\end{equation}

In order to reduce the number of parameters, we make the problem dimensionless, by using $\beta_{0}$ and $f^{-1}$ as the characteristic scales of length and time. Then after some algebra, from (\ref{eq:49}) we get
\begin{gather}
\eta_{t}^{2}=2\left(\mathfrak{H}-\mathfrak{U}\right),\label{eq:52}\\
\mathfrak{H}=\frac{1}{2}-k+\frac{H}{f^{2}\beta_{0}^{2}},
\quad\mathfrak{U}=\frac{1}{2}
\left(\eta-1\right)^{2}\left(1+\frac{k}{\eta}\right),\label{eq:53}
\end{gather}
Here the variable $\eta$ and the parameter $k$ are dimensionless quantities given as
\begin{equation}
\eta=\frac{\beta}{\beta_{0}},\quad k=\frac{2\kappa}{f^{2}\beta_{0}^{3}},\label{eq:54}
\end{equation}

\begin{figure}[t!]
\centerline{\includegraphics[width=\columnwidth]{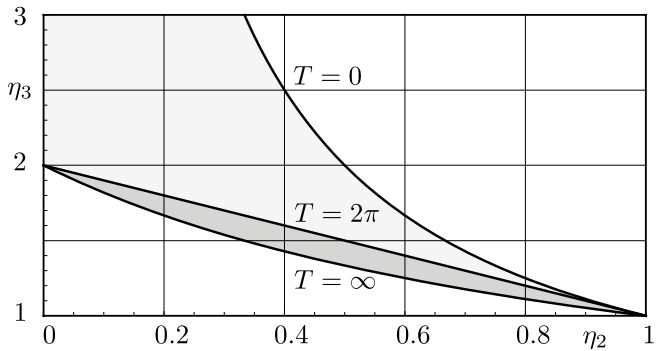}}
\caption{The domains of existence for oscillatory regimes. Their boundaries are lines of equal periods $T=0$, $T=2\pi$, $T=\infty$. The dark gray region corresponds to warm jets with $-1<k<0$, and the light gray one -- cold jets with $k>0$.}\label{fig4}
\end{figure}

Eq.~(\ref{eq:52}) can be considered as the equation of motion of a point particle in the potential $\mathfrak{U}(\eta)$. As the analysis shows (see Fig.~\ref{fig4}), the model has two oscillatory regimes, depending on the parameter $k$, whose sign coincides with the sign of the temperature deviation $\Delta T$.

\begin{figure}[t!]
\centerline{\includegraphics[width=\columnwidth]{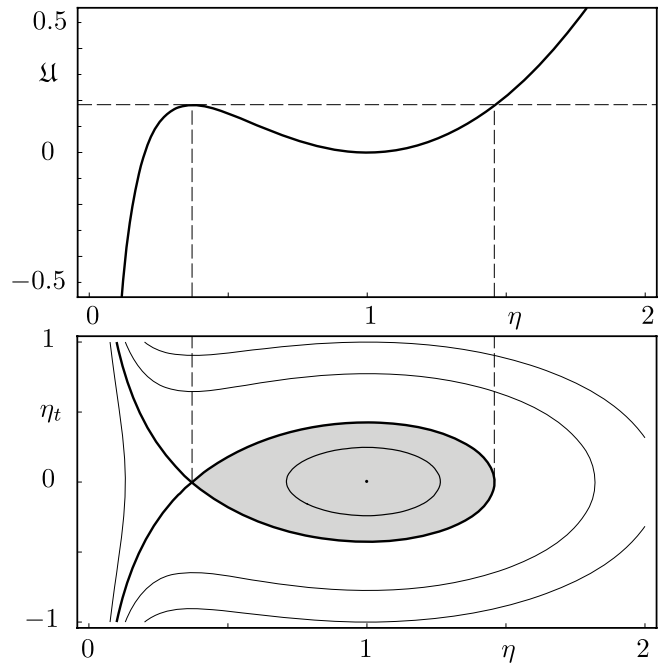}}
\caption{A sketch of the potential and phase portrait for the model in the case $\kappa<0$. The solid line marks the separatrix loop, and the oscillatory domain within it is colored gray.}\label{fig5}
\end{figure}

An oscillatory hierarchy in the theory of straight-line jets is arranged in such a way that only warm jets with $-1<k<0$ can provide long-period pulsations with $T>2\pi$. Although cold jets with $k>0$ also supports pulsations, their periods can only be less than $2\pi$. Moreover, as shown in Fig.~\ref{fig5}, in contrast to cold jets, the phase trajectories of oscillations for warm jets are surrounded by separatrices.

This fact has direct relevance to the subject-matter of this paper, since any separatrix loop implies the existence of the finite amplitudes in the system, at approaching to which the oscillation period can become arbitrarily large. For the model under consideration, in particular, this provides a mechanism through which the oscillation period due to the Earth's daily rotation can undergo significant changes under the influence of nonlinearity.

Assuming that the quantity $\eta$ oscillates in the range $\eta_{2}\leq\eta\leq\eta_{3}$, Eq.~(\ref{eq:52}) can be transformed into
\begin{equation}
\eta_{t}^{2}=\eta^{-1}\left(\eta-\eta_{1}\right)
\left(\eta-\eta_{2}\right)\left(\eta_{3}-\eta\right),\label{eq:55}
\end{equation}
where, in force of the condition $-1<k<0$, the right side roots obey to inequalities:
\begin{equation}
0<\eta_{1}<\eta_{2}<1<\eta_{3}.\label{eq:56}\\
\end{equation}

On integration of Eq.~(\ref{eq:55}) (see the Appendix for details), using the parameters $\eta_{2}$ and $\eta_{3}$ instead of $\mathfrak{H}$ and $k$ is preferable. In this case, it is necessary to have in mind that
\begin{gather}
\mathfrak{H}=\frac{1}{2}\frac{(1-\eta_{2})^{2}(1-\eta_{3})^{2}}
{1-\eta_{2}\eta_{3}},\label{eq:57}\\ k=\frac{\eta_{2}\eta_{3}(\eta_{2}+\eta_{3}-2)}{1-\eta _{2}\eta_{3}},\quad \eta_{1}=\frac{2-\eta_{2}-\eta_{3}}{1-\eta_{2}\eta_{3}}.\label{eq:58}
\end{gather}

As shown in the Appendix, by introducing the notations
\begin{equation}
p=1-\frac{\eta_{2}}{\eta_{3}},\quad m=\frac{\eta_{1}}{\eta_{2}-\eta_{1}}.\label{eq:59}
\end{equation}
the solution of (\ref{eq:55}) can be presented in parametric form
\begin{gather}
\eta=\frac{\eta_{2}}{1-p\sin^{2}\varphi},\label{eq:60}\\
t=2\sqrt{(1-p)(1+m)}\;\Pi(p;\varphi|-pm),\label{eq:61}
\end{gather}
where $\varphi$ is the phase, and $\Pi$ is the incomplete elliptic integral of the third kind.

Since the solution (\ref{eq:60}), (\ref{eq:61}) describes oscillations, their period $T$ can be found from the condition $T=t|_{\varphi=\pi/2}$, which leads us to the expression
\begin{equation}
T=4\sqrt{(1-p)(1+m)}\;\Pi(p;-pm).\label{eq:62}
\end{equation}
It should be underlined this quantity is bounded from below $T\geq2\pi$. That, in turn, means that if a nonlinear system with $-1<k<0$ is under the action of Coriolis force, inertial oscillations $T=2\pi$ should manifest themselves in it as a minimum possible period. According to (\ref{eq:59}), this is exactly the value that is reached when $p\rightarrow0$ or $m\rightarrow0$.

As may be concluded from the conditions (\ref{eq:56}), oscillatory regimes for warm jets are realized with the accordance of the inequalities
\begin{equation}
\frac{2}{1+\eta_{2}}\leq\eta_{3}\leq2-\eta_{2}.\label{eq:63}
\end{equation}
In the space of parameters $\eta_{2}$ and $\eta_{3}$, they define a narrow existence domain (see Fig.~\ref{fig4}), at whose boundaries the oscillation period $T$ and the parameter $k$ take the following values
\begin{gather}
T\Bigr|_{\eta_{3}=2-\eta_{2}}=2\pi,\quad
k\Bigr|_{\eta_{3}=2-\eta_{2}}=0,\label{eq:64}\\
T\Bigr|_{\eta_{3}=2/(1+\eta_{2})}=\infty,\quad
k\Bigr|_{\eta_{3}=2/(1+\eta_{2})}=-\frac{2\eta_{2}^{2}}
{1+\eta_{2}}.\label{eq:65}
\end{gather}

For comparison, note that for cold jets at the upper boundary of the domain, where $\eta_{3}=1/\eta_{2}$, these quantities have the following values
\begin{equation}
T\Bigr|_{\eta_{3}=1/\eta_{2}}=0,\quad
k\Bigr|_{\eta_{3}=1/\eta_{2}}=\infty.\label{eq:66}
\end{equation}

From the last Eqs.~(\ref{eq:63})--(\ref{eq:65}) we can drawn a number of conclusions. Firstly, the upper boundary in Fig.~\ref{fig4} corresponds to the simplest case of linear (harmonic) oscillations, which develop against the background of neutral thermal stratification $k=0$. Thus, in this regime, the equality
\begin{equation}
a_{+}=a_{-},\quad a_{+}=\eta_{3}-1,\quad a_{-}=1-\eta_{2}.\label{eq:67}
\end{equation}
for amplitude deviations $a_{+}$ and $a_{-}$ from the stationary state is supported exactly.

In the whole rest of the domain, the violation of this equality is quite weak. For each fixed $\eta_{2}$, the most asymmetry in oscillatory amplitudes is achieved at the lower boundary, where parameter $k$ takes the threshold value $k=-2\eta_{2}^{2}/(1+\eta_{2})$. In the point $\eta_{2}=\sqrt{2}-1$, the amplitude difference $\Delta=a_{-}-a_{+}$ reaches a peak value equal to $\Delta\approx0{.}17$.

\begin{figure}[t!]
\centerline{\includegraphics[width=\columnwidth]{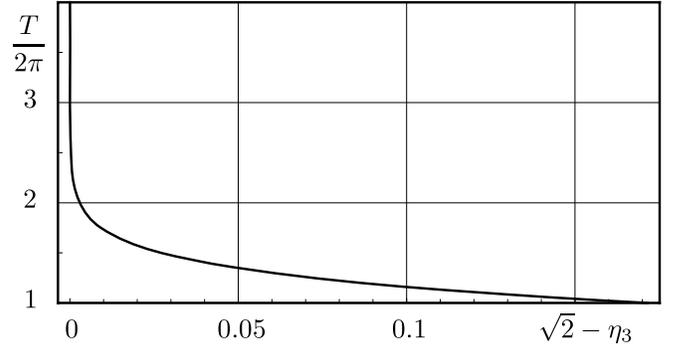}}
\caption{The period $T$ at $\eta_{2}=1-\sqrt{2}$ as a function of the deviation of the amplitude parameter $\eta_{3}$ from the threshold value $\sqrt{2}$.}\label{fig6}
\end{figure}

The only quantity that is substantially changing in the domain is the period $T$. As shown in Fig.~\ref{fig6}, this effect manifests itself as the exponential growth of $T$.
It occurs at once the amplitudes $\eta_{2}$ and $\eta_{3}$ become sufficiently close to thresholds. These values are reached on the lower boundary of the domain of existence and related one another as
\begin{equation}
\eta_{3}=\frac{2}{1+\eta_{2}}.\label{eq:68}
\end{equation}
Such behavior indicates that the phase trajectory of the oscillation is near the separatrix. As a result, the period of such oscillations can be very different from the inertial period $T=2\pi$, which is the minimum permissible for the given system.

One more scenario leading to disturbances, whose temporal durations are very large in comparison with the inertial period, occurs when their phase trajectories turn to be exactly on the separatrix. In this case the parameters $\eta_{1}$ and $\eta_{2}$ coincide, $\eta_{3}$ satisfies the equality (\ref{eq:68}), and Eq.~(\ref{eq:55}) can be transformed as
\begin{equation}
\eta_{t}^{2}=\eta^{-1}\left(\eta-\eta_{2}\right)^{2}
\left(\frac{2}{1+\eta_{2}}-\eta\right).\label{eq:69}
\end{equation}
Such solutions are called separatrix ones. They initially increase, reaching at some time the maximum value $\eta=2/(1+\eta_{2})$, and then decrease to an asymptotic level $\eta=\eta_{2}>0$.

\begin{figure}[t!]
\centerline{\includegraphics[width=\columnwidth]{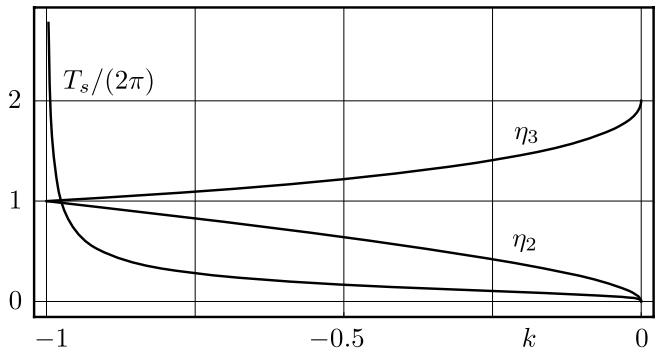}}
\caption{The time scale $T_{s}/(2\pi)$ and amplitude parameters $\eta_{2}$ an $\eta_{3}$ as a functions of $k$ for separatrix solutions.}\label{fig7}
\end{figure}

Based on this scenario, to integrate Eq.~(\ref{eq:69}) explicitly, we assume that, in the course of evolution, the system moves along the separatrix loop. In doing so let it leave the saddle point at $t=-\infty$, stop at $t=0$, and return to the saddle point at $t=+\infty$. In this case, as a result, we get
\begin{gather}
t=T_{s}\ln\left(\frac{\sqrt{\zeta}-T_{s}\sqrt{1-\zeta}}
{\sqrt{\zeta}+T_{s}\sqrt{1-\zeta}}\right) -2\arccos\sqrt{\zeta},\label{eq:70}\\
T_{s}=\sqrt{\frac{\eta_{2}(1+\eta_{2})}{(1-\eta_{2})(2+\eta_{2})}},\quad
\eta=\frac{2\zeta}{1+\eta_{2}}.\label{eq:71}
\end{gather}
The solution describes the temporal evolution during which the jet first expands and then narrows, restoring its original width at $t\rightarrow\infty$. It should be noted that the separatrix solutions have only one key parameter $k$. As shown in Fig.~\ref{fig7}, all other parameters depend on it, including $T_{s}$, which plays the role of a characteristic time.

\section{Discussion and conclusions}\label{sec6}

This paper deals with studying the dynamics of horizontal jet-streams and addresses at least two issues. The first one concerns the principal possibility of using the active layer model to achieve this goal. And the second relates to the applicability of the obtained results to some problems of geophysical fluid dynamics. The explanation of additional peaks in the macrometeorological range of the horizontal wind speed spectrum is one such problem.

The straight-line jet theory developed within the active layer model, in addition to stationary states, assumes two more dynamical possibilities for jets: transverse pulsations and rotation at a constant angular velocity. The theory also predicts that warm jets pulsate with periods longer than the period of inertial oscillations caused by the Coriolis force, while cold jets pulsate with periods shorter than inertial oscillations. So, only warm jets which satisfy the condition
\begin{equation}
-1<k<0,\label{eq:72}
\end{equation}
where $k$ defined in (\ref{eq:54}) is a dimensionless factor of the thermal stratification, can contribute to the macrometeorological spectral range.

According to the results of the theory, both jet velocity components (longitudinal one $U$ and transverse $V$) are expressed through the variable $\eta$ as follows
\begin{equation}
V=\omega\eta_{t}\frac{n}{\eta},\quad U=\omega\frac{n}{\eta} \left(\eta-1+k\right).\label{eq:73}
\end{equation}
The extremal values of these quantities are achieved at jet stream boundaries $n=\pm\beta_{0}$ and have opposite signs at them.

In the stationary state when $\eta=1$, these expressions yield the formulas
\begin{equation}
V=0,\quad U=\omega kn.\label{eq:74}
\end{equation}
which describe a uniform shear flow in a band of constant width $2\beta_{0}$. Because the profile of the longitudinal velocity $U$  is asymmetrical relative to $n=0$, such flow, in some sense, can be considered as a result of superposing two opposite-directed jets.

Depending on the parameter $\omega=\pm f$, the band with shear flow either remains immovable or rotates at the constant angular velocity $-f$. In the first case, when $\omega=f$ and hence $k<0$, the flow in the immovable band has positive vorticity
\begin{equation}
\zeta=\omega-U_{n}=\omega(1-k).\label{eq:75}
\end{equation}
In the second case, when $\omega=-f$ and the band is rotating, its flow vorticity becomes negative.

In order, relying on the jet theory, to deeper reach an insight into the investigated phenomena and correctly interpret the observable data, it is necessary to have an appropriate criterion. A suitable criterion can be derived directly from Eq.~(\ref{eq:74}) and the inequality (\ref{eq:72}) if the maximal longitudinal velocity along with the half-width $\beta_{0}$ are considered as typical values. Then this criterion will look like
\begin{equation}
|k|=\frac{U_{0}}{f\beta_{0}}<1.\label{eq:76}
\end{equation}
Here $U_{0}$ is the modulus of the maximum value of the longitudinal velocity reached at jet boundaries $n=\pm\beta_{0}$.

The criterion can be modified so that it will be able to take into account the influence of both geometric and temperature effects on jet pulsations. To this end, we need to express the stratification factor $k$ through the rest parameters of the jet. Using Eqs.~(\ref{eq:42}), (\ref{eq:5}) and applying them to axial values, we get the relationships for the jet height $h_{0}$ and the buoyancy $\tau_{0}$:
\begin{equation}
h_{0}\tau_{0}=\frac{2\kappa}{3\beta_{0}},
\quad\tau_{0}=-\gsl\frac{\Delta T_{0}}{T_{K}}.\label{eq:77}
\end{equation}
Here $\Delta T_{0}$ is the axial value of the temperature deviation from the background absolute temperature $T_{K}$.

Using (\ref{eq:77}) and (\ref{eq:54}), the stratification factor $k$ can be presented as
\begin{equation}
k=-3Bu,\quad Bu=\gsl\frac{\Delta T_{0}}{T_{K}}
\frac{h_{0}}{f^{2}\beta_{0}^{2}},\label{eq:78}
\end{equation}
where $Bu$ is the so-called Burger number.

For instance, the conditions:
\begin{gather}
g=9{.}8\,m\cdot s^{-2},\quad\beta_{0}=2\cdot 10^{5}\,m,\quad
h_{0}=10^{3}\,m,\label{eq:79}\\
f=1{.}2\cdot 10^{-4}\,s^{-1},\quad\Delta T_{0}/T_{K}=10^{-2},\label{eq:80}
\end{gather}
which are typical of the Earth's atmosphere, in accordance with (\ref{eq:78}), lead us to the equality $k=-0{.}51$. It means that in a stationary state, the longitudinal velocity of the jet takes the value $U_{0}=12{.}25\,m\cdot s^{-1}$.

\begin{figure}[htp]
\centerline{\includegraphics[width=\columnwidth]{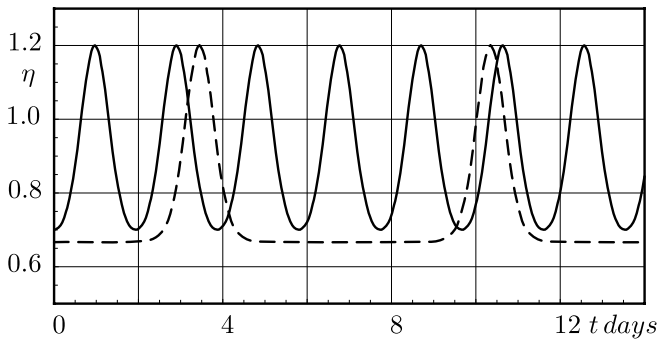}}
\caption{Synoptic jet pulsations. Two-day pulsations correspond to a solid line and seven-day pulsations to a dashed line.}\label{fig8}
\end{figure}

If the amplitude of such a jet is pulsating in the range of $\eta_{2}$ and $\eta_{3}$, these values must obey the constraint (\ref{eq:58}). In particular, if we take $\eta_{3}=1{.}2$, then at $k=-0{.}51$ we find $\eta_{2}=0{.}7$. The jet pulsations with these parameters are shown in Fig.~\ref{fig8}. Their period calculated in accordance with (\ref{eq:62}) amounts to $T=1{.}93\,days$.
\clearpage

As for the transverse velocity, it reaches its maximum at $\eta=1$, which allows us to reach the expression
\begin{equation}
V_{0}=U_{0}\frac{(1-\eta_{2})(\eta_{3}-1)}{\sqrt{1-\eta_{2}\eta_{3}}}.\label{eq:81}
\end{equation}
The calculations by this formula for the jet considered above give $V_{0}=1{.}84\,m\cdot s^{-1}$.

For comparison, the dashed line in Fig.~\ref{fig8} presents pulsations for one more jet. However, despite it has very close in amplitude parameters $\eta_{2}=0{.}6666667$ and $\eta_{3}=1{.}2$, its period $T=6{.}9\,days$ differs noticeably. It means that near the separatrix, there is a high sensitivity of the oscillation period to the jet parameters.

Obviously, in real geophysical systems, it is almost impossible that the background parameters are stationary with such high accuracy. Perhaps that is why it is quite problematic to experimentally observe regular oscillations with periods of more than 2-3 days.

\appendix
\section{}

Let's consider Eq.~(\ref{eq:55}) and enter the notations
\begin{equation}
p=1-\frac{\eta_{2}}{\eta_{3}},\quad m=\frac{\eta_{1}}{\eta_{2}-\eta_{1}}.
\label{eq:A1}
\end{equation}
The solution of Eq.~(\ref{eq:55}) will be searching in the parametric form
\begin{equation}
\eta =\frac{\eta_{2}}{1-p\sin^{2}\varphi},\quad
t=t\left(\varphi \right),\label{eq:A2}
\end{equation}
where the phase $\varphi$ plays the role of the parameter.

Then for variable $t(\varphi)$ satisfies the equation
\begin{equation}
t_{\varphi}^{2}=\frac{4(1-p)(1+m)}{\left(1-p\sin^{2}\varphi\right)^{2}
\left(1+pm\sin^{2}\varphi\right)},\label{eq:A3}
\end{equation}
which has the solution
\begin{equation}
t=2\sqrt{(1-p)(1+m)}\Pi(p;\varphi|-pm),\label{eq:A4}
\end{equation}
where $\Pi$ is the incomplete elliptic integral of the third kind.

Since the positive parameters $\eta_{1}$, $\eta_{2}$, and $\eta_{3}$ are arranged as $\eta_{1}<\eta_{2}\leq\eta_{3}$, the parameters $p$ and $m$ satisfy to inequalities
\begin{equation}
0\leq p<1,\quad m\geq0.\label{eq:A5}
\end{equation}

\acknowledgments
This work was equally supported by the Russian Foundation for Basic Research (Project No.~18-05-00831), by the Presidium of the Russian Academy of Sciences (Program of Nonlinear Dynamics in Mathematical and Physical Sciences), and by the Russian Science Foundation (Project No.~19-17-00248).



\begin{thebibliography}{99}

\bibitem{rhi94}
P.\,B. Rhines,
``Jets,''
Chaos \textbf{4}, 313--339 (1994).

\bibitem{val06}
G.\,K. Vallis,
{\sl Atmospheric and Oceanic Fluid Dynamics: Fundamentals and Large-scale Circulation},
(Cambridge University Press, 2006).

\bibitem{hh13}
J. Holton and G. Hakim,
{\sl An Introduction to Dynamic Meteorology},
Academic Press (Elsevier Science, 2013).

\bibitem{vdh57}
I. Van der Hoven,
``Power spectrum of horizontal wind speed in the frequency range from 0.0007 to 9000 cycles per hour,''
Journal of Meteorology \textbf{14}, 160--164 (1957).

\bibitem{kw15}
S. Kang and H. Won,
``Intra-farm wind speed variability observed by nacelle anemometers in a large inland wind farm,''
J. Wind Eng. Ind. Aerodyn. \textbf{147}, 164--175 (2015).

\bibitem{kw16}
S. Kang and H. Won,
``Spectral structure of 5 year time series of horizontal wind speed at the Boulder Atmospheric Observatory,''
J. Geophys. Res. Atmos. \textbf{121}, 946--967 (2016).

\bibitem{del03}
P.\,J. Dellar,
``Common Hamiltonian structure of the shallow water equations with horizontal temperature gradients and magnetic fields,''
Physics of Fluids \textbf{15}, 292--297 (2003).

\bibitem{gp12}
V.\,P. Goncharov and V.\,I. Pavlov,
``Blow-up instability in shallow water flows with horizontally-non-uniform density,''
JETP Lett. \textbf{96}, 427--431 (2012).

\bibitem{gp13a}
V.\,P. Goncharov and V.\,I. Pavlov,
``Structural elements of collapses in shallow water flows with horizontally nonuniform density,''
JETP \textbf{117}, 754--763 (2013).

\bibitem{gp13b}
V.\,P. Goncharov and V.\,I. Pavlov,
``Simple model of the Rayleigh-Taylor instability, collapse, and structural elements,''
Phys. Rev. E \textbf{88}, 023002 (2013).

\bibitem{gp15}
V.\,P. Goncharov and V.\,I. Pavlov,
 ``Algebraic instability in shallow water flows with horizontally nonuniform density,''
Phys. Rev. E \textbf{91}, 043004 (2015).

\bibitem{gp16}
V.\,P. Goncharov and V.\,I. Pavlov,
``Pulsating jet-like structures in magnetized plasma,''
Physics of Plasmas \textbf{23}, 082117 (2016).

\bibitem{knnr07}
E.\,A. Kuznetsov, V. Naulin, A.\,H. Nielsen, and J.\,J. Rasmussen,
``Effects of sharp vorticity gradients in two-dimensional hydrodynamic turbulence,''
Physics of Fluids \textbf{19}, 105110 (2007).

\bibitem{gp14}
V.\,P. Goncharov and V.\,I. Pavlov,
``Whether the Sun's supergranulation possesses a scaling?''
JETP Lett. \textbf{99}, 317 (2014).

\bibitem{rip93}
P. Ripa,
``Conservation laws for primitive equations models with inhomogeneous layers,''
Geophys. Astrophys. Fluid Dynam. \textbf{70}, 85--111 (1993).

\bibitem{ztl18}
V. Zeitlin,
{\sl Geophysical Fluid Dynamics: Understanding (almost)
Everything with Rotating Shallow Water Models},
(Oxford University Press, 2018).

\bibitem{and84}
D.\,T.\,T. Anderson,
``On equatorial dynamics, mixed layer physics and sea-surface temperature,''
Tellus, Ser. A \textbf{36}, 278--291 (1984).

\bibitem{rip99}
P. Ripa,
``On the validity of layered models of ocean dynamics and thermodynamics with reduced vertical resolution,''
Dynam. Atmos. Oceans \textbf{29}, 1--40 (1999).

\end{thebibliography}

\end{document}